\documentstyle[aps,prl]{revtex}

\begin{document}

\twocolumn[\hsize\textwidth\columnwidth\hsize\csname
@twocolumnfalse\endcsname

\title{Hole dynamics in noble metals}

\author{I. Campillo$^{1}$, A. Rubio$^{2}$, J. M. Pitarke$^{1,5}$, A.
Goldmann$^3$, and P. M. Echenique$^{4,5}$}

\address{$^1$Materia Kondentsatuaren Fisika Saila, Zientzi Fakultatea,
Euskal Herriko Unibertsitatea,\\ 644 Posta kutxatila, 48080 Bilbo, Basque
Country, Spain\\
$^2$Departamento de F{\'{\i}}sica Te\'orica, Universidad de Valladolid,
47011 Valladolid, Spain\\
$^3$Fachbereich Physik, Universit\"at Gh Kassel, Heinrich-Plett-Strabe 40,
D-34132 Kassel, Germany\\
$^4$Materialen Fisika Saila, Kimika Fakultatea, Euskal Herriko
Unibertsitatea,\\
1072 Posta kutxatila, 20080 Donostia, Basque Country, Spain\\
$^5$Donostia International Physics Center (DIPC) and Centro Mixto
CSIC-UPV/EHU,\\
Donostia, Basque Country, Spain\\
}

\date\today

\maketitle

\begin{abstract}
We present a detailed analysis of hole dynamics in noble metals (Cu and Au), by
means of first-principles many-body calculations. While holes in a free-electron
gas are known to live shorter than electrons with the same excitation energy,
our results indicate that $d$-holes in noble metals exhibit longer inelastic
lifetimes than excited $sp$-electrons, in agreement with experiment. The
density of states available for $d$-hole decay is larger than that for the decay
of excited electrons; however, the small overlap between $d$- and $sp$-states
below the Fermi level increases the $d$-hole lifetime. The impact of $d$-hole
dynamics on electron-hole correlation effects, which are of relevance in the
analysis of time-resolved two-photon photoemission experiments, is also
addressed.\\
\end{abstract}

]

Detailed and quantitative understanding of hot-electron and photo-hole dynamics
is the prerequisite for any tailoring of technological processes in solid-state
physics and surface chemistry, which are governed by charge transfer and
electronic excitations\cite{Ogawa0,Hofer,Diau,Bonn}. In particular, the
quasiparicle lifetime, i.e., the time a quasiparticle propagates without loosing
its phase memory, represents a basic quantity in the analysis of these
processes. Experimentally, angle-resolved photoemission (PE) spectroscopy
provides direct information on hole
lifetimes\cite{Matzdorf,Matzdorf2,Goldm,Purdie}. A new path for the study of
both electron and hole (and spin) dynamics in the time domain was opened by the
advent of the time-resolved two-photon photoemission (TR-TPPE)
technique\cite{Bokor,Haight}. In these "pump-probe" experiments, the emitted
photoelectron is measured both with energy and momentum resolution; electron
and hole lifetimes are then measured at well-defined ${\bf k}$-space points, by
combining the band mapping capabilities of photoemission with the time
resolution of nonlinear optical spectroscopy\cite{Petek0}.

The theoretical framework to investigate the quasiparticle lifetime has been
based for many years on the free-electron gas (FEG) model of Fermi
liquids\cite{review}, characterized by the electron-density parameter
$r_s$\cite{rs}. In this simple model and for either electrons or holes with
energy $E$ very near the Fermi level ($E\sim E_F$), the inelastic lifetime
is found to be, in the high-density limit ($r_s<<1$),
$\tau(E)=263\,r_s^{-5/2}\,(E-E_F)^{-2}\,{\rm fs}$, where $E$ and
$E_F$ are expressed in ${\rm eV}$\cite{QF}. Several other free-electron
calculations of electron-electron scattering rates have also been carried out,
for electron/hole energies that deviate from the Fermi level, within the
random-phase approximation (RPA)\cite{Ritchie,Shelton} and with inclusion of
exchange and correlation effects\cite{Ashley,Penn}. Nevertheless, detailed
TR-TPPE experiments have reported large deviations of measured hot-electron
lifetimes from those predicted within the FEG
model\cite{Fann,Schmu,Hertel,Aes,Ogawa,Cao,Knoesel1}. Moreover, while holes
($E<E_F$) in a FEG are known to live shorter than their electron counterparts
($E>E_F$) [for the same excitation energy, the momentum of the hole is smaller
than the electron momentum, thus yielding a larger number of available
transitions], recent TR-2PPE measurements\cite{Petekh} have
demonstrated clearly that the $d$-hole lifetime at the top of the Cu $d$-bands is
considerably slower than that of excited electrons with the same excitation
energy. Subsequently, recent PE measurements\cite{Gold2} have arrived at the same
conclusion for all three noble metals Cu, Ag, and Au. First-principles
calculations of electron lifetimes in a variety of metals have been carried out
only very recently\cite{Igorprl,Ekardt}, by using the GW approximation of
many-body theory\cite{Hedin}. These calculations\cite{comment} show that
band-structure effects play a key role in the quasiparticle-decay mechanism.
Furthermore, hot-electron lifetimes are found to strongly depend on the momentum
of the quasiparticle, especially in the case of metals with a band structure
having van Hove-like singularities in the vicinity  of the Fermi
level\cite{Igorprl}. 

During the electron-pump process induced by the first ultrashort laser in
TR-TPPE experiments, a hole is created in the Fermi sea. This hole
can be filled with an electron from below the Fermi level, via
Auger decay, higher-order excitations, or electron-phonon interactions. 
The understanding of the
hole decay is fundamental for the description of the ulterior dynamics of
the excited electron (probed by the second laser).
Furthermore, the dynamics of the hole is interesting by itself, as
it is a fingerprint of differences between electron and hole wave functions.
These differences yield distinct behaviours of electron and hole lifetimes,
which depend on both the energy and the momentum of the quasiparticle.

In the present letter, we address the first stage of a TR-2PPE
experiment by investigating the hole-quasiparticle dynamics, and report
first-principles many-body calculations of the decay rate of $d$-like holes in Cu
and Au. Details of the formalism we use for the evaluation of decay rates are
described in Ref.\onlinecite{Igorprl}. The basic equation for the damping rate of
a hole in the state $\phi_{n,{\bf k}}({\bf r})$ with energy
$\varepsilon_{n,{\bf k}}<E_F$ is (we use atomic units throughout, i.e.,
$e^2=\hbar=m_e=1$)
\begin{equation}\label{eq1}
\tau^{-1}_{n,{\bf k}}=-2\int d{\bf r}\int d{\bf r}'\,\phi_{n,{\bf
k}}^*({\bf r})\,{\rm Im}\,\Sigma({\bf r},{\bf r}';\varepsilon_{n,{\bf
k}})\,\phi_{n,{\bf k}}({\bf r}'),
\end{equation}
where $\Sigma({\bf r},{\bf r}';\varepsilon_{n,{\bf k}})$ represents
the quasiparticle self-energy, which we compute within the GW approximation of
many-body theory\cite{Hedin}. For the evaluation of both the initial
state of Eq. (\ref{eq1}) and all wave functions entering the quasiparticle
self-energy, we first expand the one-electron Bloch states in a plane-wave
basis (PW), and then solve self-consistently the Kohn-Sham
equation\cite{Kohn} of density-functional theory (DFT). Though
all-electron schemes, such as the full-potential linearized augmented
plane-wave (LAPW) method\cite{lapw}, are expected to be better suited for the
description of $d$-bands, the PW pseudopotential approach has already
been succesfully incorporated in the description of the dynamical response of
copper\cite{Igorcu}. Both PW and LAPW calculations produce almost identical
results, showing the correct overall band structure. Nevertheless,
they both predict the entire $d$-band manifold to be $\sim 0.5\,{\rm eV}$
higher than shown by photoemission experiments\cite{Strokov,threshold}. 

The quasiparticle decay rate in periodic crystals depends on both the wave
vector ${\bf k}$ and the band index $n$ of the initial Bloch state.
Nevertheless, we also define $\tau^{-1}(E)$, as the average of
$\tau^{-1}({\bf k},n)$ over all wave vectors and bands lying with the
same energy $E$ in the irreducible wedge of the BZ. Our full
band-structure calculations of the average lifetime $\tau(E)$ of holes ($E<E_F$)
in Cu and Au are presented by open circles in Figs. 1a and 1b,
respectively\cite{details,Z}. For comparison, full band-structure calculations of
the average lifetime of electrons ($E>E_F$) are exhibited by solid circles, and
FEG calculations [with $r_s=2.67$ for Cu and $r_s=3.01$ for Au] are represented
by solid ($E>E_F$) and dotted ($E<E_F$) lines. These results indicate that holes
in noble metals exhibit considerably longer lifetimes than electrons with the
same excitation energy, which is in agreement with accurate
TR-2PPE and PE measurements of $d$-hole lifetimes reported in
Refs.\onlinecite{Petekh} and \onlinecite{Gold2}, respectively. Moreover, at
energies near the top of the $d$-bands the hole lifetime strongly deviates from
the $\tau(E)\propto(E-E_F)^{-2}$ quadratic dependence predicted within the FEG
model, as first pointed out by Petek {\it et al}\cite{Petekh}.

Both electrons and holes exhibit lifetimes that are well over those predicted
within the FEG model of the solid, due to a major contribution from occupied
$d$ states participating in the screening of electron-electron interactions,
and differences between electron and hole lifetimes stem from the intrinsic
properties of Bloch states above and below the Fermi level. Whereas Bloch states
above the Fermi level have mainly $sp$-like character, states just below the
Fermi level have a small but significant $d$-component, which
increases when the $d$-band threshold is reached. This difference accounts for
the larger hole lifetime for energies very near the Fermi level, as shown in
Fig. 1. At the top of the $d$-bands [$1.5$ and $1.7\,{\rm eV}$ below
the Fermi level in Cu and Au, respectively], the low overlap between $d$- and
$sp$-states below the Fermi level yields a dramatic increase in the hole
lifetime, especially in the case of Cu, which cannot be explained with use of
the FEG model. As the hole energy decreases ($|E-E_F|$ increases), both the
increased phase space for the hole to decay and the larger overlap for hole-hole
scattering within the $d$-bands lead to a rapid decrease of the hole
lifetime\cite{Petekh}.

This physical scenario for the hole dynamics in noble metals is drawn in Fig. 2,
where the ratio between hole and electron lifetimes in Cu (solid circles) and Au
(open circles) is plotted as a function of the hole/electron energy with respect
to the Fermi level. Though calculated lifetimes of both electrons and holes in Au
are much longer than in Cu, due to a larger participation of occupied
$d$ states in the dynamical screening of the former\cite{lifeau}, different
ratios between hole and electron lifetimes are only observed at energies near
the top of the $d$ bands. As $5d$-bands in Au are more
free-electron-like than $3d$-bands in Cu, the larger overlap between $d$ and
$sp$-states below the Fermi level in Au yields an increase in the hole lifetime
at $|E-E_F|\sim 1.7\,{\rm eV}$ in this material that is not as dramatic as in
the case of Cu. This different behaviour of hole lifetimes in Cu and Au
emphasizes the crucial role that the overlap of the $d$-hole with unoccupied
$sp$- and $d$-hole states play in the hole decay mechanism. As $|E-E_F|$
increases, the ratio between hole and electron lifetimes resembles that
predicted by the FEG model of the solid, which is plotted in the inset of Fig. 2
for $r_s=2.67$.

$d$-bands in the noble metals open in the vicinity of the high-symmetry
$X$-point. Hence, we have calculated lifetimes of holes with
the wave vector along the $\Gamma X$ direction in Cu. At the top of
the $d$-bands, at the $X_5$ point with $E-E_F=-1.5\,{\rm eV}$ and the $X_2$
point with $E-E_F=-1.7$, we obtain $d$-hole lifetimes of $99$ and $90\,{\rm
fs}$, respectively, much larger than lifetimes of electrons with the same
energy, showing the role that the low overlap of
$d$-holes with unoccupied $sp$-hole states play in the determination of the hole
lifetime. Though these calculated values of $d$-hole lifetimes cannot be
directly compared with experiment, since the measured $d$-band threshold
in Cu is located at $\sim 2\,{\rm eV}$ below the Fermi
level, our calculations are in agreement with the
experimental observation that the narrowest linewidths correspond to the top of
the $d$-bands\cite{Petekh,Gold2}. In the case of holes with the wave vector along
the $\Gamma L$ direction in Cu, the hump of Fig. 1a at $E-E_F\sim-1.5\,{\rm eV}$
is absent and the calculated lifetimes decrease with a quadratic
[$\tau\propto(E-E_F)^{-2}$] energy scaling. Instead, the hump of Fig. 1a is
mainly originated in the contribution to the average lifetime from ${\bf k}$ vectors 
along the $\Gamma X$ direction, where the opening of the $d$-bands occurs,
thereby showing a distinct
behaviour of hole lifetimes along various symmetry directions\cite{Peteknew}.

As $d$-holes in noble metals are found to live longer than $sp$-electrons
with the same excitation energy, the majority of electrons excited by the
first probe pulse in a TR-2PPE experiment feels the field created by the existing
hole, thereby altering the electron dynamics. Hence, the TR-2PPE hot-electron
lifetime measurements include, in a complex way, contributions from the joint
electron-hole (exciton) dynamics\cite{Gumhalter}. In the case of semiconductors,
this excitonic renormalization is known to strongly modify the single-particle
optical absorption profile, so it certainly needs to be included in
electronic-screening calculations\cite{excitons}. In
principle, excitonic normalization does not play such an important role in the
case of metals, because of the large dynamical screening in these materials.
However, there is a femtosecond time scale for the building up of the screening.
During this ultrashort time, the electron-hole interaction is not fully screened
and might modify the lifetime of the excited electron, as measured by the second
probe pulse. Hence, there is still much to be done to understand the
time-dependent building up of the screening in real metals, as well as the
many-body electron and hole correlation effects in the quasiparticle dynamics.

In summary, we have presented a detailed theoretical investigation of
inelastic lifetimes of holes in Cu and Au. We have reached the important
conclusion that $d$-holes in these materials exhibit longer lifetimes than
$sp$-electrons with the same excitation energy, in agreement with experiment.
While a major contribution from occupied $d$-states participating in the
screening of electron-electron interactions yields both electron and hole
lifetimes that are much longer than those of electrons and holes in a FEG, the
small overlap between $d$- and
$sp$-states below the Fermi level is responsible for the lifetime of
$d$-holes being longer than that of $sp$-electrons, especially at the top of the $d$-band. As the $d$-hole energy
decreases ($|E-E_F|$ increases), both the increased density of states available
for the hole-decay mechanism and the larger overlap for hole-hole scattering
within the $d$-bands yield a rapid decrease of the hole lifetime. Although our
full band-structure calculations predict longer lifetimes than measured in the
experiment\cite{new}, they are in agreement with the experimental observation that the
narrowest linewidths correspond to the top of the $d$-bands. Also,
our results highlight new effects related to the electron-hole interaction
occurring during the time delay between the two laser pulses in TR-2PPE
spectroscopy, which may be of great importance in the interpretation of these
experiments. The present results are general, they can be applied to the study
of other $d$-metals, and can also be extended to the investigation of spin
dynamics.

We acknowledge partial support by the University of the Basque
Country, the Basque Hezkuntza, Unibertsitate eta Ikerketa Saila,
and the Spanish Ministerio de Educaci\'on y Cultura.

\begin{figure}
\caption[]{Electron and hole lifetimes in (a) Cu and (b) Au. Solid and open
circles represent our full {\em ab initio} calculation of $\tau(E)$ for
electrons ($E>E_F$) and holes ($E<E_F$), respectively, as obtained after
averaging $\tau({\bf k},n)^{-1}$ of Eq. (\ref{eq1}) over wave vectors and the 
band structure for each ${\bf k}$. The solid and dotted lines
represent the corresponding lifetime of electrons (solid line) and holes (dotted
line) in a FEG with $r_s=2.67$ for Cu and $r_s=3.01$ for Au.}
\end{figure}

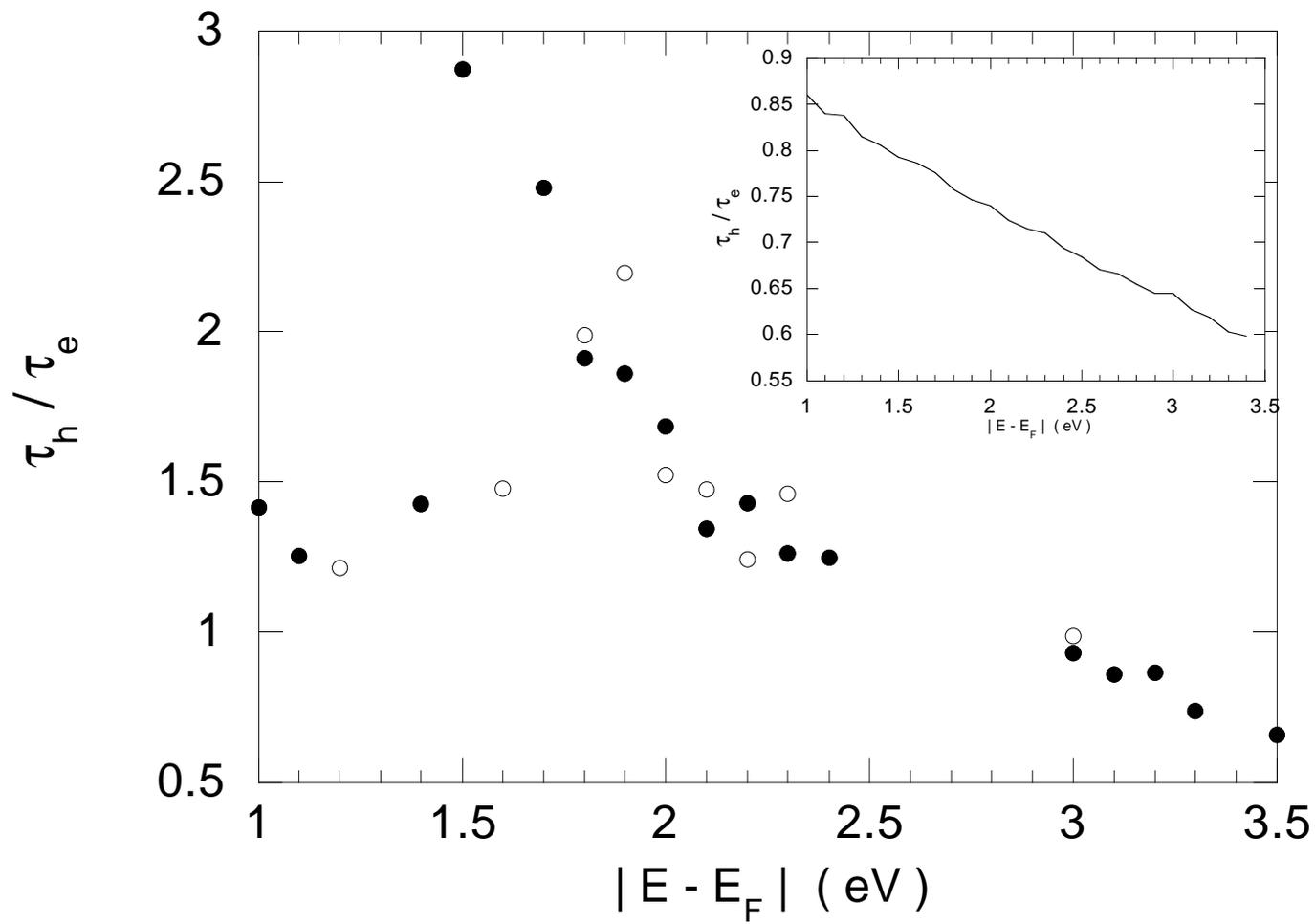
\begin{figure}
\caption[]{Ratio between hole and electron lifetimes in Cu (solid circles) and
Au (open circles). The inset exhibits the corresponding ratio between hole and
electron lifetimes in a FEG with $r_s=2.67$.}
\end{figure}


\begin{references}

\bibitem{Ogawa0} S. Ogawa {\it et al}, Phys. Rev. Lett. {\bf 78}, 1339 (1997).
\bibitem{Hofer}  U. H\"ofer {\it et al}, Science {\bf 277}, 1480 (1997).
\bibitem{Diau} E. W. G. Diau {\it et al}, Science {\bf 279}, 847 (1998).
\bibitem{Bonn}  M. Bonn {\it et al}, Science {\bf 285}, 1042 (1999).
\bibitem{Matzdorf} R. Matzdorf {\it et al}, Solid
State Commun. {\bf 92}, 839 (1994).
\bibitem{Matzdorf2} R. Matzdorf, Appl. Phys. A {\bf 63}, 549 (1996); Surf. Sci.
Rep. {\bf 30}, 153 (1998).
\bibitem{Goldm} A. Goldmann, R. Matzdorf, and F. Theilmann, Surf. Sci.
{\bf 414}, L932 (1998).
\bibitem{Purdie} D. Purdie {\it et al}, Surf. Sci. {\bf 407}, L671 (1998).
\bibitem{Bokor} J. Bokor, Science {\bf 246}, 1130 (1989)
\bibitem{Haight} R. Haight, Surf. Sci. Rep. {\bf 21}, 275 (1995).
\bibitem{Petek0} H. Petek and S. Ogawa, Prog. Surf. Sci. {\bf 56}, 239 (1997).
\bibitem{review} P. M. Echenique {\it et al}, Chem. Phys {\bf 251}, 1 (2000); and
references therein.
\bibitem{rs} $r_s$ is defined by the relation
$1/n_0=(4/3)\pi r_s^3$, with $n_0$ the average electron density.
\bibitem{QF} J. J. Quinn and R. A. Ferrell, Phys. Rev. {\bf 112}, 812
(1958).
\bibitem{Ritchie} R. H. Ritchie, Phys. Rev. {\bf 114}, 644 (1959)
\bibitem{Shelton} J. J. Quinn, Phys. Rev. {\bf 126}, 1453 (1962); J. C. Shelton, 
Surf. Sci. {\bf 44}, 305 (1974).
\bibitem{Ashley} R. H. Ritchie and J. C. Ashley, J. Phys. Chem. Solids {\bf
26}, 1689 (1963); L. Kleinman, Phys. Rev. B {\bf 3}, 2982 (1976); D. R. Penn, 
Phys. Rev. B {\bf 13}, 5248 (1976).
\bibitem{Penn} D. R. Penn, Phys. Rev. B {\bf 22}, 2677 (1980).
\bibitem{Fann} W. S. Fann {\it et al}, Phys. Rev. Lett. {\bf 68}, 2834 (1992);
Phys. Rev. B {\bf 46}, 13592 (1992).
\bibitem{Schmu} C. A. Schmutenmaer {\it et al}, Phys. Rev.
B {\bf 50}, 8957 (1994).
\bibitem{Hertel} T. Hertel {\it et al}, Phys. Rev. Lett. {\bf 76}, 535 (1996).
\bibitem{Aes} M. Aeschlimann {\it et al}, Phys. Rev. Lett. {\bf 79}, 5158
(1997).
\bibitem{Ogawa} S. Ogawa, H. Nagano, and H. Petek, Phys. Rev. B {\bf 55}, 1
(1997).
\bibitem{Cao} J. Cao {\it et al}, Phys. Rev. B {\bf 56}, 1099 (1997); Phys. Rev.
B {\bf 58}, 10948 (1998).
\bibitem{Knoesel1} E. Knoesel, A. Hotzel, and M. Wolf, Phys. Rev. B {\bf 57},
12812 (1998).
\bibitem{Petekh} H. Petek, H. Nagano, and S. Ogawa, Phys. Rev. Lett.
{\bf 83}, 832 (1999).
\bibitem{Gold2} R. Matzdorf {\it et al}, Appl. Phys. B {\bf 68}, 393 (1999).
\bibitem{Igorprl} I. Campillo {\it et al}, Phys. Rev. Lett. {\bf 83}, 2230
(1999); I. Campillo {\it et al}, Phys. Rev. B {\bf 61}, 13484 (2000).
\bibitem{Ekardt} W.-D. Sch\"one {\it et al}, Phys. Rev. B {\bf 60}, 8616
(1999); R. Keyling, W.-D. Sch\"one, and W. Ekardt, Phys. Rev. B {\bf 61}, 1670
(2000).
\bibitem{Hedin} L. Hedin and S. Lundqvist, Solid State Phys. {\bf 23}, 1 (1969).
\bibitem{comment} In the experiments there are various physical processes
contributing to the measured lifetime that are not
taken into account in these calculations. These are electron
transport of the excited electron, higher-order electron-hole
recombination processes, and electron-phonon interactions [which might be
important and even dominant for high enough temperatures and
very-low-energy electrons].
\bibitem{Kohn} P. Hohenberg and W. Kohn, Phys. Rev. {\bf 136}, B864
(1964); W. Kohn and L. Sham, Phys. Rev. {\bf 140}, A1133 (1965).
\bibitem{lapw} D. J. Singh, {\it Plane Waves, Pseudopotentials, and the LAPW
method} (Kluwer, Boston, 1994).
\bibitem{Igorcu} I. Campillo, A. Rubio, and J. M. Pitarke, Phys. Rev.
B {\bf 59}, 12188 (1999).
\bibitem{Strokov} V. N. Strokov {\it et al}, Phys. Rev. Lett. {\bf 81}, 4943
(1998).
\bibitem{threshold} The computed $d$-band energies do not
improve when gradient corrections to the LDA exchange-correlation potential
are introduced. The existing disagreement between theory and experiment is
expected to be due to self-energy corrections to exchange and correlation.
\bibitem{details} Well-converged results have been found with the introduction
of a kinetic-energy cutoff of 75 Ry, thereby accounting for all
$3d^{10}$ and $5d^{10}$ electrons of Cu and Au, respectively. The electron-ion
interaction has been described by means of a non-local, norm conserving 
Troullier-Martins pseudopotential. The dielectric matrix has been evaluated in
the random-phase approximation (RPA), by including bands up to a maximum energy
of $25\,{\rm eV}$ above the Fermi level. The sampling over the Brillouin zone
(BZ) has been performed on a $16\times\ 16\times 16$ Monkhorst-Pack mesh. The
exchange-correlation potential has been evaluated within the local-density
approximation (LDA), by using the Perdew-Zunger parametrization of the
Ceperley-Alder correlation energy. Crystalline local-field effects are fully
included in our calculations.
\bibitem{Z} These calculations have been carried out by computing the
self-energy on the energy-shell, with no explicit inclusion of the
excitation-spectral-weight renormalization that is due to changes of the
self-energy near the Fermi level. Within a FEG model of the solid this
renormalization is known to yield lifetimes that are {\it longer} than those
obtained on the energy-shell by $\sim 20\%$ (see Ref.\onlinecite{review}).
\bibitem{lifeau} I. Campillo {\it et al}, Phys. Rev. B {\bf 62}, 1500 (2000). 
\bibitem{Peteknew} H. Petek, H. Nagano, M. Weida, and S. Ogawa, Chem. Phys.
{\bf 251}, 71 (2000).
\bibitem{Gumhalter} B. Gumhalter and H. Petek, Surf. Sci. {\bf 445}, 195
(2000). 
\bibitem{excitons} M. Rohfling and S. G. Louie, Phys. Rev. Lett. {\bf
81}, 2312 (1998); L. X. Benedict, E. L. Shirley, and R. B. Bohn,
Phys. Rev. Lett. {\bf 80}, 4514 (1998); S. Albretcht {\it et al}, Phys. Rev.
Lett. {\bf 80}, 4510 (1998).
\bibitem{new} The Auger recombination rate reported in Ref.\onlinecite{Petekh}
was determined by substracting the electron-phonon contribution from the total
decay rate. However, it might very well be that the $T=0$ rate contribution from
the electron-phonon coupling is larger than estimated in Ref.\onlinecite{Petekh},
and the measured value of $\tau(E)$ should, therefore, be considered as a lower 
limit of the
actual inelastic lifetime.
\end{references}
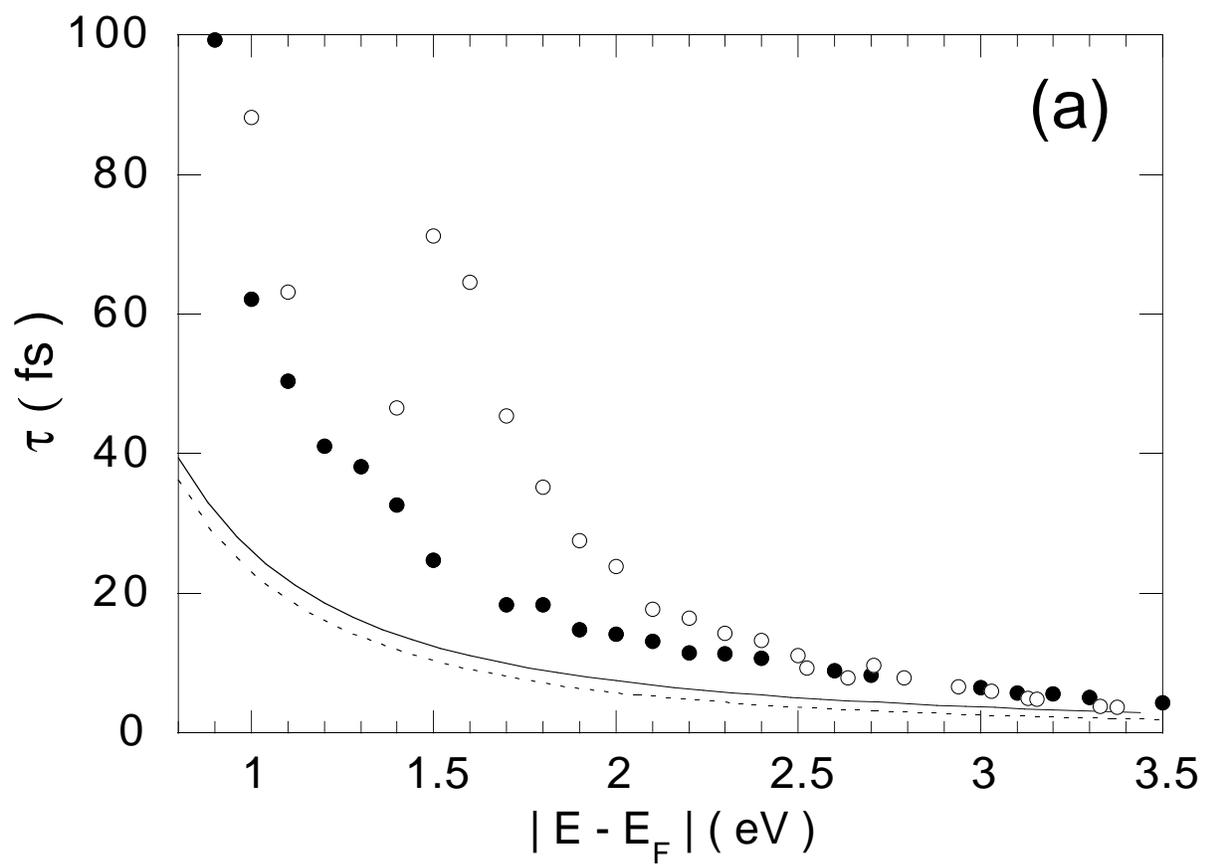
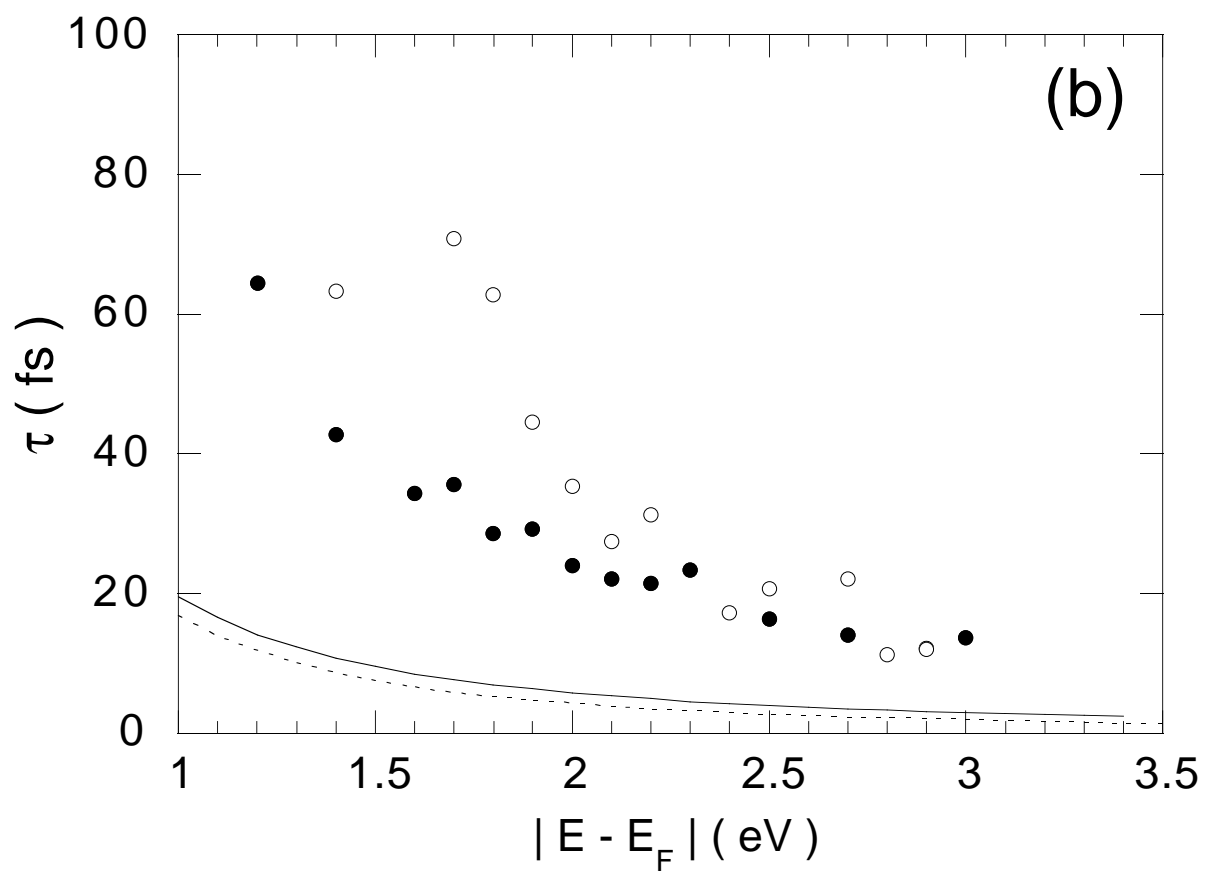
\end{document}